# Preliminary Study on Bit-String Modelling of Opinion Formation in Complex Networks

Yi Yu, Gaoxi Xiao
School of Electrical and Electronic Engineering,
Nanyang Technological University,
Singapore.
e-mail: egxxiao@ntu.edu.sg

*Abstract*—Opinion formation has been gaining increasing research interests recently, and various models have been proposed. These models, however, have their limitations, among which noticeably include (i) it is generally assumed that adjacent nodes holding similar opinions will further reduce their difference in between, while adjacent nodes holding significantly different opinions would either do nothing, or cut the link in between them; (ii) opinion mutation, which describes "opinion changes not due to neighborhood influences" in real life, is typically random. While such models enjoy their simplicity and nevertheless help reveal lots of useful insights, they lack the capability of describing many complex behaviors which we may easily observe in real life. In this paper, we propose a new bit-string modeling approach. Preliminary study on the new model demonstrates its great potentials in revealing complex behaviors of social opinion evolution and formation.

## I. INTRODUCTION

Opinion propagation, evolution and formation play a critical role in shaping our society and influences almost every aspect of our life, from as "small" as interpersonal relationship [1] to as big as elections [2][3], etc. There are several works on the propagation of different opinions in social networks [4]-[6] and the impacts of opinion propagation on the social structures [7][8], etc. Another important topic is how people's opinions are influenced by each other in their social interactions and how such opinion changes help shape the opinion groups. Such is known as *opinion formation* problem.

Extensive studies have been conducted on opinion formation in social population and a few different models have been proposed [9]-[22]. The simplest one among them is probably the voter model [10]-[12]. It assumes that there are only two opinions in the population, representing positive and negative attitudes towards a certain incident, respectively. In every time step, a randomly selected node (or an individual in the network; hereafter "individual" and "node" shall be used interchangeably) may adopt the opinion of its randomly selected neighbor. The voter model has been extended to the case with multiple different opinions [13][14]. Other works typically quantify the opinion as continuous variable [15]-[20]. Two most well-studied models include bounded confidence model [15]-[17] and the Deffuant model [18]-[22]. Both models assume that a node's opinion can be influenced by those neighbors who hold similar, or at least not-so-different, opinions, termed as *similar opinion neighbors* (SONs) hereafter. The only difference is that while Deffuant assumes that a node's opinion may be affected by a randomly selected SON, the bounded confidence model assumes that all the SONs have combined influences on the node. In both models, there is *consensus making*, while the node's opinion and its randomly selected SON (or all SONs) come closer to each other. Note that, in both models two opinions are regarded as similar opinions if the difference between them is smaller than a given *tolerance* value $d$. For the Defffuant model, existing results show that the network would enter into a final state where several opinion groups are formed and coexist. The number of groups has a linear relationship with $1/d$.

*Noise* was first introduced in Deffuant model in [20]-[22] to simulate the change of views for any reasons other than a SON's influence. In these studies, it was assumed that all opinions have an equal chance to change to any other opinion (In the rest of this letter, we term such change as *mutation*.). The results showed that the final-state opinion distribution shall resemble a well-defined bell curve [20] and the initial conditions have hardly any effects on the final state, with the only exception of some very special cases (e.g., the initial opinion is of a single value in the whole system) [21].

Limitation of adopting such a simple mutation model in the Deffuant model was revealed in [23]. It was shown that when different opinions have different chances of having mutations, the system dynamics may become rather complex. In fact, for different distributions of the "mutation probability" within the range of opinion, different final steady state may be achieved. In that study, however, it was still assumed that once a mutation happens, the "target" of the mutation is randomly distributed; in other words, the opinion may change to any other opinion with an equal chance.

We may argue that opinion mutation in the real life may not have a randomly distributed target in most cases. Everyone is "defined" and "bounded" by his/her current and/or historical states to a certain extent. Some mutations may be relatively easier to happen than the others. In other words, for each opinion the mutation target may also have a non-uniform distribution; and more importantly, different opinions may have different non-uniform target opinion distributions. In other words, the distribution of mutation target may rely on its current (or even historical) state. A new modeling approach capable of revealing such kind of state dependent mutation is in demand. The random target opinion distribution commonly adopted in current literature shall be viewed as a special case of the requested new modeling approach, where the distribution of the mutation target is independent of a node's current or historical state.

With the understanding of the limitation that the conventional opinion mutation models may have, it would be interesting to also have a look at the conventional consensus making models from this new angle as well. We may realize that the conventional consensus making model is state dependent: whether two neighbors could make consensus depends on the opinions they are holding. While such is applausible, the way that similar opinions come closer to each other may be more complex than what this model can describe. For example, people making consensus may stick to some of their differences, if such differences matter to them: close friends may tend to agree on almost everything, except for one or two "small but important" issues. What may be even more important is that, people with significantly different ideas may have very different chances of cutting the link in between them, depending on what that or those significant differences are.

To make an effort towards tackling the shortcomings of the conventional models as discussed above, in this paper, inspired by the genetic mutation in nature [24], we propose a new bit-string modeling approach. Specifically, we use a string of binary numbers to represent an opinion or a set of opinions. By doing so, we may (i) reflect the importance/relevance of different opinions or different part of an opinion where a higher bit represents a more important/relevant opinion among a set of opinions held by the individual, or a more important part of an opinion held by the individual; and (ii) conveniently reflect the different mutation target distributions of different opinions or different part of an opinion, e.g., by assigning different bits with different probabilities of mutation. It would not be difficult to take one step further by assigning "0" and "1" at different bit positions with different probabilities of mutation, reflecting the case where the probabilities of opinion change in two opposite directions are not symmetric. Our preliminary studies show that such an approach may have great potentials to reveal the complex dynamics of opinion formation in social networks which cannot be conveniently revealed by any of the existing models to the best of our knowledge.

The rest of this letter is organized as follows. Section 2 briefly describes the Deffuant model and then introduces the bit-string opinion model. As a case study, Section 3 discusses on a simple case where the mutation probabilities from 0 to 1 and from 1 to 0 are different on each bit position. We will see that the simple case nevertheless leads to some interesting and complex behaviors. Section 4 concludes the letter.

## II. MODEL DESCRIPTION

### A. Review of Deffuant Model with Mutation

Deffuant model assumes that opinions are continuously distributed within the interval [0, 1]. At each time step $t$, a node $A$ is randomly selected together its random neighbor $B$. Denote their opinions as $o(t, A)$ and $o(t, B)$, respectively. If the difference between these two opinions is less than a given tolerance $d$, they make consensus according to the following rules:

$$\begin{cases} o(t+1, A) = o(t, A) - \mu[o(t, A) - o(t, B)]; \\ o(t+1, B) = o(t, B) + \mu[o(t, A) - o(t, B)]. \end{cases} \quad (1)$$

A smaller value of $\mu$ may slow down the evolution process while different values of $\mu$, as long as it is within the range of (0, 1/2], is believed to lead to the same final steady state [18]. Hereafter, we use $\mu = 1/2$ as that in most of the existing works.

Noise/mutation was firstly introduced into Deffuant model in [20]. Specifically, in each time step $t$, a randomly selected node has a probability $p$ to mutate and adopt another randomly chosen opinion.

### B. Bit-string modelling approach

The bit-string model is based on a simple idea of describing an opinion or a set of opinions into a string of binary number. For example, an opinion, or a set of opinions, adopted by an individual in a certain circumstance may be written as 01101001. Higher bits may denote something that is more "fundamental" and important to an individual, e.g., whether s/he has any religion belief in a study on "opinion formation of people's interpretation of eternity in a social community", while a lower bit may be generally speaking less significant, e.g., the individual's preference of sport activities in the above study. Certainly a string can also be used to represent a single idea (e.g., the religion belief in the above example), while different bits are of different importance in defining the idea: 01101001 may be regarded as a similar idea to 01101010, but significantly different from 11101001. In the above example, the former case means that two individuals have nearly the same religion belief in almost every detail; while in the latter one, the two individuals are very different in their religion beliefs.

At the first sight, adopting a bit-string model may be of limited benefits: it would be the same thing to write 01101001 as 105 in decimal number, or 105/255 as a real number within the range of [0,1]. The benefits, however, lie in the convenience of defining different "behaviors" on different bits. For example, by defining different bits with different mutation probabilities, we may resemble the fact that changing an individual's religion belief may be easier or more difficult than changing his/her favorite sports activities, both of which may affect, in rather different ways, how likely or unlikely his/her social connections can change his/her interpretation of eternity. Further, for the bit corresponding to religion belief of the individual, assigning different probabilities for it to change from 0 to 1 and change from 1 to 0 respectively would resemble the real-life case that it is easier or more difficult to make free thinker become a religion believer, or go through the opposite direction. The potentials of such a new modeling approach are attractive.

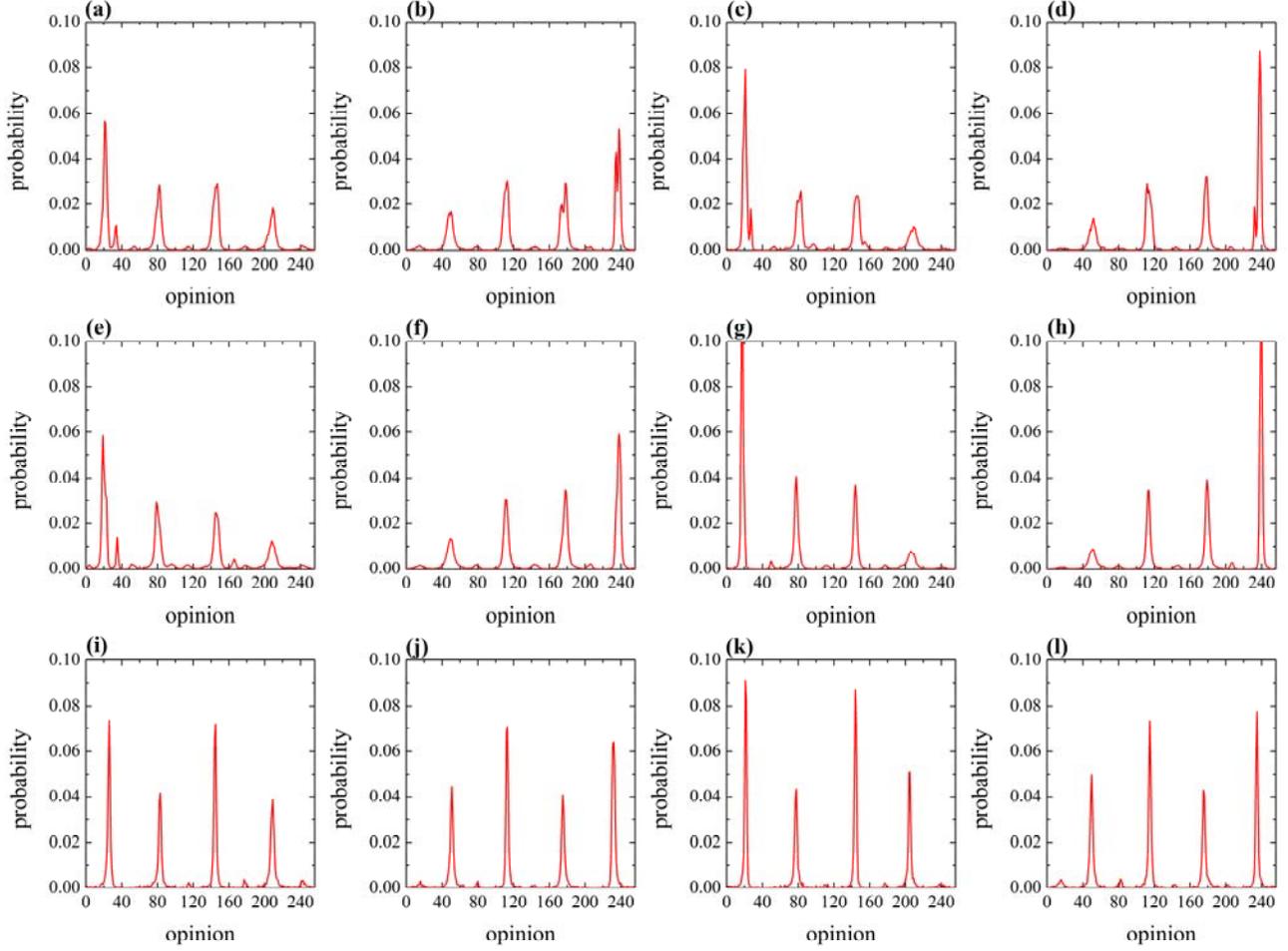

Figure 1. Final opinion distribution at $d = 25$ for different $(p_{10}, p_{01})$ *i)* when each bit of the string has the same probability of mutation: (a) (0.01, 0.005), (b) (0.005, 0.01), (c) (0.01, 0.003), (d) (0.003, 0.01); *ii)* when $\alpha = 1/28$ hence higher bits have higher probabilities of mutation.: (e) (0.01, 0.005), (f) (0.005, 0.01), (g) (0.01, 0.003), (h) (0.003, 0.01); and *iii)* when $\alpha = -1/28$ hence lower bits have higher probabilities of mutation: (i) (0.01, 0.005), (j) (0.005, 0.01), (k) (0.01, 0.003), (l) (0.003, 0.01).

There are many different ways to define how different opinions may interact with each other and mutate themselves by using this bit-string model. For example, it would not be difficult to imagine crossover between two bit strings, like that in the genetic algorithm [25]. In this preliminary study, we consider the simple case which essentially is still the well-known Deffuant model with mutation, with the only difference the *i*-th bit has a mutation probability $p(i)$ which may be different for different bit positions (i.e., different values of *i*). While $p(i)$ may be affected by various combinations of many different current/historical factors as we discussed earlier, we consider the simple case where $p(i)$ is only affected by the current state of the *i*-th bit. Specifically, we consider the case where $p(i)$ is composed of two conditional probabilities: the probability for the *i*-th bit to mutate from 1 to 0 given that its current state is 1, and the probability of mutating from 0 to 1 given that its current state is 0, denoted as $p_{10}(i)$ and $p_{01}(i)$ respectively. Apparently, we have

$$p(i) = p_{01}(i)q_0(i) + p_{10}(i)q_1(i), \qquad (2)$$

where $q_0(i)$ and $q_1(i)$ denote the probabilities that the current state of the *i*-th bit is 0 and 1, respectively. Note that, in the above model, since $q_0(i)$ and $q_1(i)$ evolve with the network system, $p(i)$ is time varying until the system reaches steady state. This is very different from that in the existing studies where the mutation rate is typically a constant throughout the evolution process. We argue, however, that in the real life, mutation rate may be indeed time varying in most cases: a system in transition is expected to witness a relatively higher mutation rate, which may become lower when the system enters into a relatively more stable state.

Also note that (1) does not necessarily lead to an integer value that can be written into a binary bit string, in which case we assign the closest integer opinion to the node, and a tie is broken arbitrarily.

### III. SIMULATION RESULTS AND DISCUSSIONS

We simulate the simple case where each opinion is represented by an 8-bit string (or equivalent 0-255 in decimal number). In each time step, in addition to the standard consensus making operation as that in the conventional Deffuant model, a node will be randomly selected as the candidate of opinion mutation. For the selected node, a single bit will be selected as the bit with a non-zero probability of having a mutation, where the $i$-th bit of the opinion is selected at a probability $\rho(i)$, $\sum_{i=1}^{8}\rho(i)=1$. We consider the case where the same set of values of $p_{10}(i)$ and $p_{01}(i)$ apply to all the network nodes and all the 8 bits. Specifically, we examine 4 pairs of different $p_{10}(i)$ and $p_{01}(i)$: (0.01, 0.005), (0.005, 0.01), (0.01, 0.003) and (0.003, 0.01), respectively. Note that by adopting such small values, a bit mutation does not happen more frequently than an average of 1 in every 100 time steps. We present the results in the ER random network [26] with a size of $N=10^4$ and an average nodal degree of $z=10$.

We start by considering the case where $\rho(i)=1/8, i=1,2,...8$. Setting the tolerance $d=25$, we perform the simulation for $t=5\times 10^7$ time steps for each case and average the opinion distribution of the last 1000 steps as the final-state opinion distribution. Figures 1(a) to 1(d) illustrate the final state for the four different cases respectively. From Figures 1(a) and 1(c), the observation is that when $p_{10}>p_{01}$, the peaks positioned at smaller values would be higher; meanwhile the positions of the four peaks also slightly shift to the left-hand side. When $p_{10}<p_{01}$, the observations we can make from Figures 1(b) and 1(d) go to the opposite: the peaks positioned at larger values are higher and the peak positions shift to the right. The differences between the heights of different peaks become larger when the ratio between $p_{10}$ and $p_{01}$ is larger in the former case (comparing Figures 1(a) and 1(c)) and smaller in the latter case (comparing Figures 1(b) and 1(d)). Such observations match our daily experiences. For example, when the whole society tends to be optimistic (pessimistic), though people may still holding different ideas, different ideas may all tend to be shifted towards the optimistic (pessimistic) side. The more optimistic (pessimistic) the society is, the more people would be found at the optimistic (pessimistic) end, and the peaks of opinions typically also shift to that end. Though such observations are well known in real life, to the best of our knowledge, it is the first time that it is observed in numerical simulation based on a simple mathematical model.

We then consider the slightly different case that

$$\rho(i)=\alpha(4.5-i)+0.125, \quad i=1,2,\ldots 8 \qquad (3)$$

where $\alpha \in [-1/28, 1/28]$. For this function, a positive $\alpha$ means that higher order digits have higher probabilities to be selected for mutation while a negative $\alpha$ indicts the opposite. We still set $d=25$.

Figures 1(e) to (h) and Figures 1(i) to (l) present the results when $\alpha=1/28$ and $-1/28$, respectively. Note that when $\alpha=1/28$, higher order bits have higher probabilities to be selected for mutation, at a ratio of $\rho(i):\rho(j)=(j-1):(i-1), i,j=1,2,\ldots,8;$ while for $\alpha=-1/28$, lower order bits have higher probabilities to be selected, and the ratio becomes $\rho(i):\rho(j)=(i-1):(j-1)$, $i,j=1,2,\ldots,8$.

For $\alpha=1/28$, Figures 1(e) to 1(h) respectively present the final opinion distributions corresponding to 4 pairs of different $p_{10}(i)$ and $p_{01}(i)$: (0.01, 0.005), (0.005, 0.01), (0.01, 0.003) and (0.003, 0.01). The observations are almost the same as those in Figures 1(a) to 1(d). The only nontrivial difference is that in Fig. 1(g) (1(h)), the peak at the leftmost (rightmost) side is much higher than the corresponding peak in Fig. 1(c) (1(d)). A rough understanding of the reasons behind is not so difficult to achieve: when $p_{10}(i)$ is much higher $p_{01}(i)$ and higher bits have higher chances of mutation, the chance of having "0" on higher bits becomes higher, making the peaks closer to the left side end higher. This explains the observation in Figures 1(g). Similar reasoning can be adopted to explain the difference between Figures 1(d) and 1(h). Considering that Figures 1(a) and 1(e) however appear to be nearly the same, it remains as a challenge to figure out how big a difference between $p_{10}(i)$ and $p_{01}(i)$ is big enough to lead to nontrivial differences in the final state.

Figures 1(i) to 1(l), however, present very different observations when lower bits have higher probabilities to mutate: while peaks still shift to the left when $p_{10}>p_{01}$ (Figures 1(i) and (k)) and to the right when $p_{10}<p_{01}$ (Figures 1(j) and (l)), the heights of the four peaks do not increase or decrease monotonically from left to right. Rough understanding may still be easily achieved: when higher bits have lower opportunities of mutation and the highest bit has a zero mutation probability (and therefore does not mutate at all), at steady state we shall expect to find half of the nodes holding opinions starting with a bit "0" and the other half a bit "1". Opinion distribution is thus roughly 50-50 on the left and right half of the opinion axis. Mutation of the other bits

(2nd to the 8th bits) can still generate "uneven" distribution in each half of the opinion axis, depending on whether $p_{10} > p_{01}$ or $p_{10} < p_{01}$.

While rough understandings as discussed above are not difficult to achieve, obviously extensive further studies are needed to fully understand the system dynamics.

## IV. CONCLUSION

In this letter, we proposed a new bit-string modeling approach for more efficiently describing the complex dynamics of opinion formation in complex networks. The new approach allows convenient modeling of various non-uniform, state-dependent behaviors of different opinions or different parts of an opinion. Preliminary study on a very simple case reveals the great potentials the new approach may have.

A lot of other interesting observations have been made in our preliminary studies, which have been largely omitted in this letter. These observations shall be carefully sorted into some systematic descriptions and discussions in our future studies. A theoretical framework for analyzing the evolution of the system adopting the new modeling approach will also be developed.

ACKNOWLEDGMENT

This work is partially supported by Ministry of Education (MOE), Singapore, under research grant RG 28/14 and MOE2013-T2-2-006.